\documentclass[a4paper,11pt]{article}
\usepackage[DIV12]{typearea}
\usepackage{longtable}
\usepackage{booktabs,colortbl}
\usepackage{multirow}
\usepackage{amsmath}
\usepackage{amssymb}
\usepackage{bbm}
\usepackage[utf8]{inputenc}
\usepackage{pdflscape}
\usepackage{xspace}
\usepackage[caption=false]{subfig}
\usepackage{nicefrac}
\usepackage{graphicx}
\usepackage{cite}  
\usepackage[small]{caption2}
\usepackage{scalerel} 
\usepackage{mathtools}

\usepackage{adjustbox}
\usepackage{slashed}
\usepackage{authblk}

\numberwithin{equation}{section}

\newcommand{\ttitle}{Brick wall diagrams as a completely integrable system}

\newcommand{\e}{\mathrm{e}}
\newcommand{\I}{\mathrm{i}}

\newcommand{\dd}{\mathrm{d}}
\newcommand{\tr}{\mathrm{tr}}
\newcommand{\Tr}{\mathrm{Tr}}

\usepackage{xcolor}
\definecolor{darkgreen}{HTML}{109930}
\definecolor{pink}{rgb}{0.858, 0.188, 0.478}

\advance \headheight by 3.0truept       
\usepackage{hyperref}
\hypersetup{
    pdftitle = {\ttitle},
    pdfauthor = {Kade, Staudacher}
}

\begin{document}

\begin{titlepage}

\begin{flushright}
\normalsize{HU-EP-23/52-RTG}\\
\end{flushright}

\vspace*{1.0cm}

\begin{center}
{\Large\textbf{\boldmath \ttitle}\unboldmath}

\vspace{1cm}

\textbf{Moritz Kade}, \textbf{Matthias Staudacher}
\\[4mm]
\begin{small}
\texttt{\string{\href{mailto:mkade@physik.hu-berlin.de}{mkade},\href{mailto:staudacher@physik.hu-berlin.de}{staudacher}\string}@physik.hu-berlin.de}
\end{small}
\\[8mm]
\textit{\small Institut f\"{u}r Mathematik und Institut f\"{u}r Physik,\\
Humboldt-Universit\"{a}t zu Berlin,\\
Zum Großen Windkanal 2, 12489 Berlin, Germany}
\end{center}

\vspace{1cm}

\vspace*{1.0cm}

\begin{abstract}
We study the free energy of an integrable, planar, chiral and non-unitary four-dimensional Yukawa theory, the bi-fermion fishnet theory discovered by Pittelli and Preti. The typical Feynman-diagrams of this model are of regular ``brick-wall''-type, replacing the regular square lattices of standard fishnet theory. We adapt A.\ B.\ Zamolodchikov's powerful classic computation of the thermodynamic free energy of fishnet graphs to the brick-wall case in a transparent fashion, and find the result in closed form. Finally, we briefly discuss two further candidate integrable models in three and six dimensions related to the brick wall model.
\end{abstract}

\end{titlepage}

\newpage

\section{Introduction and results}
\label{sec:IntroductionAndResults}

The discovery and use of integrability of quantum field theories (QFT) is often strongly related to techniques initially established for exactly solvable models of statistical physics, see the prime example of planar $\mathcal{N}=4$ super Yang-Mills theory (SYM) \cite{Beisert:2010jr}. 
Concerning QFTs in more than two dimensions, A.\ B.\ Zamolodchikov showed in 1980 \cite{Zamolodchikov:1980mb} that massless propagators of lattice-like vacuum Feynman diagrams akin to ``fishing-net'' diagrams can be interpreted as certain weights analogous to Boltzmann weights appearing in integrable lattice models. More concretely, a star-triangle 
relation \cite{DEramo:1971hnd,UniquenessDIKazakov}, sometimes also called method of uniqueness,
for the propagators results in a Yang-Baxter equation, thereby leading to the manifest appearance of quantum integrability.
To demonstrate the power of his observation, Zamolodchikov calculated the free energy of his system in the thermodynamic limit. It corresponds to the critical coupling of a then still unknown QFT, producing the asymptotic growth of the latter's vacuum graphs. To be more precise, he applied the method of inversion relations \cite{StroganovInvRelLatticeModels,Baxter:1982xp,Baxter:1989ct,Baxter:2002qg,Pokrovsky_1982,bousquetmelou1999inversion}, which had been developed for integrable statistical models, to the QFT's propagator-weights.

Many years later (2015) the conformal, non-unitary QFT that actually produces the examined graphs (and essentially nothing else!) in the limit $\mathrm{N}\rightarrow\infty$ was constructed by Gürdo\u{g}an and Kazakov \cite{Gurdogan:2015csr} and christened ``bi-scalar fishnet theory''. 
In fact, it turns out to be a special case of an even more general class of such theories obtained by performing a double-scaling limit of $\gamma$-deformed planar $\mathcal{N}=4$ SYM \cite{Beisert:2005if}, yielding the so-called $\chi$-CFT \cite{Gurdogan:2015csr} with three effective couplings corresponding to the Cartan elements of the R-symmetry algebra $\mathfrak{so}(6)\cong\mathfrak{su}(4)$ \cite{Caetano:2016ydc}. Here all of the gauge fields of $\mathcal{N}=4$ SYM disappear, along with one of its four fermions. Setting two out of these three effective couplings to zero yields the special case of Zamolodchikov's bi-scalar fishnet theory. 
The authors of \cite{Gurdogan:2015csr} also briefly discussed an intermediate version of these models named $\chi_0$-CFT that still contained all of the three scalars of the $\chi$-CFT and two out of its three fermions. A further simplified version of this ``intermediate'' model, called bi-fermion fishnet theory by them, was then proposed by Pittelli and Preti \cite{Pittelli:2019ceq}. It contains only one scalar, but keeps the two fermions and will be the focus of the present paper. We will call this theory {\it brick wall model}, according to the shape of its Feynman diagrams. This name was proposed for the these types of diagrams in \cite{Chicherin:2017frs}.
Extending and adapting Zamolodchikov's  approach to a larger class of diagrams including fermions, we show that the asymptotic scaling of the model's vacuum graphs can once again be determined by a known star-triangle relation \cite{Chicherin:2012yn} and calculate its critical coupling by deriving and solving suitable inversion relations. Interestingly, while the brick wall model cannot be obtained by a double-scaling limit of planar $\gamma$-deformed $\mathcal{N}=4$ SYM, the authors of \cite{Pittelli:2019ceq} derived it from a strong, imaginary Lunin-Maldacena deformation of $\mathcal{N}=2$ superconformal field theory (SCFT). 

Along with Zamolodchikov's free energy, other observables such as certain anomalous dimensions and correlation functions \cite{Gromov:2018hut} could be determined exactly for the bi-scalar fishnet model, which may be interpreted as a large-$\mathrm{N}$ (planar) matrix field toy model version of four-dimensional $\phi^4$-theory. Unlike its textbook relative, it is, albeit non-unitary, integrable \cite{Gromov:2017cja} and conformal \cite{Sieg:2016vap,Grabner:2017pgm}.
Moreover, increasing the complexity of bi-scalar fishnet theory back to the direction of full-fledged planar $\mathcal{N}=4$ SYM showed that integrability continues to provide exact answers for anomalous dimensions and correlators \cite{Kazakov:2018gcy,Chicherin:2017frs,Derkachov:2020zvv}. More precisely, 
the $\chi_0$- and $\chi$-models stayed integrable and conformal even after including fermions with their characteristic Yukawa interactions. Some of this work was also adapted to the brick wall model, see \cite{Pittelli:2019ceq}, which may be interpreted as a large-N (planar) matrix field toy model version of four-dimensional Yukawa theory.

Our paper is organized as follows: In section \ref{sec:DerivationAndMotivationOfTheBrickwallModel} we discuss the Lagrangian of the brick wall model, explain its graphical motivation from non-planar $\mathcal{N}=4$ SYM, and very briefly discuss its quantization. In section \ref{sec:STRForScalarsAndFermions} we review the star-triangle relation (STR) and its consequences for the manipulation of Feynman integrals using integrability. 
Section \ref{sec:VacuumGraphsInTheThermodynamicLimit} is devoted to the vacuum graphs in the large-$\mathrm{N}$ limit and to their description by generalized graph building operators in similarity to partition functions of statistical models. 
Section \ref{sec:InversionRelationsOfTheFreeEnergyByIntegrabilty} then proceeds with the calculation of the critical coupling by the method of inversion relations. Finally section \ref{sec:Outlook} gives an outlook on future applications of the brick wall theory and exhibits two related candidate integrable models in three and six dimensions.
There are two appendices: appendix~\ref{app:DerivationOfXUnity} proves the main relation used to derive the inversion relations. In appendix~\ref{app:DerivationOfScalarResults} we transparently review Zamolodchikov's famous but extremely concise results \cite{Zamolodchikov:1980mb} for the asymptotics of vacuum fishnet graphs in $D=3,4,6$ dimensions. 

\section{Graphical motivation of the brick wall model}
\label{sec:DerivationAndMotivationOfTheBrickwallModel}


The existence of the brick wall model, apart from its derivation in\cite{Pittelli:2019ceq}, is in some graphical sense also motivated by the relation betweeen the $\chi$-CFT to fishnet theory. We start by recalling the derivation of the former from the $\mathcal{N}=4$ model \cite{Gurdogan:2015csr}.
As a first step, one performs a three-parameter $\gamma$-deformation of $\mathcal{N}=4$ by modifying the product between two fields $A\cdot B$ to a star-product
\begin{equation}
A\cdot B 
\rightarrow 
\e^{-\frac{\I}{2}\, \mathrm{det}\left( \mathbf{q}_A \vert \mathbf{q}_B \vert \mathbf{\gamma}\right)}\, 
A\cdot B \;,
\end{equation}
where $\mathbf{q}_A$ and $\mathbf{q}_B$ are the $\mathfrak{su}(4)$ $R$-symmetry weight vectors of $A$ and $B$. This introduces factors $q_i := \e^{-\frac{\I}{2}\gamma_i}$ into the $\mathcal{N}=4$ Lagrangian, where $i=1, 2, 3$.
Secondly, the double-scaling limit consists of taking the 't Hooft coupling $\lambda \rightarrow 0$ and the twist parameters\footnote{A more general double-scaling procedure, where each $q_i$ is allowed to either go to zero or to infinity, was suggested in \cite{Caetano:2016ydc,Ahn:2020zly}. However, up to redefining couplings and permuting fields, it does not lead to any new models.}
$q_i \rightarrow \infty$  (by considering imaginary angles $\gamma_i \rightarrow \I\infty$). The two limits are balanced such that $\xi_i := q_i \cdot \lambda$ is finite. 
Note that the limit taking renders the theory non-hermitian, with only chiral\footnote{Chiral, not in the usual particle physics sense, but meaning that each vertex defines the same orientation on the curve each graph in the genus expansion is embedded into.}
interaction vertices remaining.
The new finite, effective coupling constants $\xi_i$ can now be tuned to describe different models. In the case of generic three parameters $\xi_i$, this theory has been christened $\chi$-CFT \cite{Gurdogan:2015csr,Caetano:2016ydc,Kazakov:2018gcy}  by Gürdo\u{g}an and Kazakov, with the Lagrangian 
\begin{equation}
\begin{split}
&\mathcal{L}^{\chi} 
~=~
\mathrm{N}  \cdot\mathrm{tr}\left[
\sum_{j=1}^3 
\left(
- \frac{1}{2} \partial^\mu \phi_j^\dagger \partial_\mu \phi_j^{\phantom{\dagger}}
+ \I \bar{\psi}_j \slashed{\partial} \psi_{j}
\right)
+\xi_1^2 \phi_2^\dagger \phi_3^\dagger \phi_2^{\phantom{\dagger}} \phi_3^{\phantom{\dagger}} 
+\xi_2^2 \phi_3^\dagger \phi_1^\dagger \phi_3^{\phantom{\dagger}} \phi_1^{\phantom{\dagger}} 
+\xi_3^2 \phi_1^\dagger \phi_2^\dagger \phi_1^{\phantom{\dagger}} \phi_2^{\phantom{\dagger}} \right.
\\
&\Biggl. 
+\I \sqrt{\xi_2 \xi_3} 
\left(
\psi_3\phi_1 \psi_2 + \bar{\psi}_3^{\phantom{\dagger}} \phi_1^\dagger \bar{\psi}_2^{\phantom{\dagger}}
\right)
+\I \sqrt{\xi_1 \xi_3} 
\left(
\psi_1\phi_2 \psi_3 + \bar{\psi}_1^{\phantom{\dagger}} \phi_2^\dagger \bar{\psi}_3^{\phantom{\dagger}}
\right)
+\I \sqrt{\xi_1 \xi_2} 
\left(
\psi_2\phi_3 \psi_1 + \bar{\psi}_2^{\phantom{\dagger}} \phi_3^\dagger \bar{\psi}_1^{\phantom{\dagger}}
\right)
\Biggr] .
\end{split}
\label{eq:LagrangianChi}
\end{equation}
All fields here and in the following transform in the adjoint representation of the ``gauge group'' $\mathrm{SU}(\mathrm{N})$, which has become a global symmetry: The gauge field interactions of $\mathcal{N}=4$ are suppressed in the limit, and we may thus completely project out the gauge fields. As is common, we omit all the Lorentz indices related to the fermions, e.\ g.\ $\psi_{i}^{\alpha}\psi_{j,\alpha}$ and $\bar{\psi}_{i,\dot{\alpha}}\bar{\psi}_{j}^{\dot{\alpha}}$ become, respectively, $\psi_i \psi_j$ and $\bar{\psi}_i\bar{\psi}_j$, and so on. The spacetime derivative $\partial_\mu$ is contracted with the Weyl matrices $\sigma_\mu$. Our convention in Euclidean space is $\sigma_\mu = (-\I \vec{\boldsymbol{\sigma}}, \mathbbm{1}_2)$, such that $\slashed{\partial}_{\dot{\alpha}}^{\alpha} = (\sigma^\mu )_{\dot{\alpha}}^{\alpha} \partial_\mu$, where $\sigma_i$ are the Pauli matrices. 
Actually, \eqref{eq:LagrangianChi} it is not quite complete, since quantum corrections generate a number of double-trace terms required to render the model fully conformal\cite{Fokken:2013aea,Sieg:2016vap}. These however only appear in a very restricted class of quantities, typically those that involve length-two operators. This is also true for all the models in the remainder of this chapter; we will only explicitly exhibit the double-trace terms for the brick wall model to follow below, cf.\ \eqref{eq:doubletraceBW}.

Thirdly, as an intermediate step, still more general than bi-scalar fishnet theory, we obtain the $\chi_0$-CFT, see \cite{Caetano:2016ydc,Kazakov:2018gcy}, by setting $\xi_1=0$. The theory has the Lagrangian 
\begin{equation}
\begin{split}
\mathcal{L}^{\chi_0} 
~=~
\mathrm{N}  \cdot\mathrm{tr}\Biggl[\Biggr. & 
- \frac{1}{2} \sum_{j=1}^3  \partial^\mu \phi_j^\dagger \partial_\mu \phi_j^{\phantom{\dagger}}
+ \sum_{k=2}^3 \I \bar{\psi}_k \slashed{\partial} \psi_{k} 
 \\
&\Biggl.
+\xi_2^2 \phi_3^\dagger \phi_1^\dagger \phi_3^{\phantom{\dagger}} \phi_1^{\phantom{\dagger}}
+\xi_3^2 \phi_1^\dagger \phi_2^\dagger \phi_1^{\phantom{\dagger}} \phi_2^{\phantom{\dagger}}
+\I \sqrt{\xi_2 \xi_3} 
\left(
\psi_3\phi_1 \psi_2 + \bar{\psi}_3^{\phantom{\dagger}}\phi_1^\dagger \bar{\psi}_2^{\phantom{\dagger}}
\right)
\Biggr] .
\end{split}
\label{eq:LagrangianChi0}
\end{equation}
Finally, to obtain the bi-scalar fishnet theory (FN), one puts also $\xi_2=0$ and defines for simplicity $\xi := \xi_3$. This results in
\begin{equation}
\begin{split}
&\mathcal{L}^{\mathrm{FN}} 
~=~
\mathrm{N}  \cdot \mathrm{tr}\left[ 
\frac{1}{2}\sum_{j=1}^2 \partial^\mu \phi_j^\dagger \partial_\mu \phi_j^{\phantom{\dagger}}
+ \xi^2 \phi_1^\dagger \phi_2^\dagger \phi_1^{\phantom{\dagger}} \phi_2^{\phantom{\dagger}}
\right] .
\end{split}
\label{eq:LagrangianFN}
\end{equation}
Now the theory contains only one chiral interaction term, which in the limit $\mathrm{N}\rightarrow\infty$ generates fishnet-like Feynman diagrams. The more general $\chi_0$-CFT \eqref{eq:LagrangianChi0} contains two vertices of this type, alongside two Yukawa vertices that couple fermions and bosons. 

We now contemplate exclusively focussing on the Yukawa interaction vertices, {\it ad-hoc} entirely eliminating the fields $\phi_2$ and $\phi_3$ from \eqref{eq:LagrangianChi0}. This results in the Lagrangian of the 
{\it brick wall model}, first obtained from a double-scaling limit of $\mathcal{N}=2$ SCFT  in \cite{Pittelli:2019ceq}:
\begin{equation}
\begin{split}
&\mathcal{L}^{\mathrm{BW}} 
~=~
\mathrm{N}  \cdot \mathrm{tr}\left[ 
- \frac{1}{2} \partial^\mu \phi^\dagger \partial_\mu \phi 
+ \I \sum_{k=1}^2 \bar{\psi}_k \slashed{\partial} \psi_{k}
+ 
\rho 
\cdot 
\left(
\psi_1\phi \psi_2 + \bar{\psi}_1\phi^\dagger \bar{\psi}_2
\right)
\right] .
\end{split}
\label{eq:LagrangianBW}
\end{equation}
It is important to note that the brick wall model {\it cannot} be obtained by a special choice of the parameters $\xi_i$ from \eqref{eq:LagrangianChi0} or \eqref{eq:LagrangianChi}. 
Its vertices are still chiral, since the hermitian conjugates of the Yukawa interaction terms, for which the fields are ordered in reverse, are missing.  This chirality, in combination with the fact that the two vertices have to alternate when following a fermion line, result in the typical ``brick wall'' structure of the generated Feynman diagrams, see figure \ref{fig:BWExample} below. 
A similar set of fishnet-inspired integrable QFTs, the so-called loom models, have recently been proposed in \cite{Kazakov:2022dbd}.

As already mentioned above, just like the $\chi$-CFT model,
the brick wall model \eqref{eq:LagrangianBW} requires a double-trace (dt) counter-term for full consistency as a conformal field theory \cite{Pittelli:2019ceq}.
It may be derived just like for $\mathcal{L}^{\chi}$ in \eqref{eq:LagrangianChi}, cf.\ \cite{Fokken:2013aea,Kazakov:2018gcy}, and reads
\begin{equation}
\mathcal{L}^{\mathrm{BW}}_\mathrm{dt} 
~=~
\alpha\cdot
\mathrm{tr}
\bigl[ 
\phi\phi
\bigr]
\mathrm{tr}
\bigl[ 
\phi^\dagger\phi^\dagger
\bigr] .
\label{eq:doubletraceBW}
\end{equation}
In order to reach the conformal fixpoint, the parameter $\alpha$ has to be finetuned as a suitable function $\alpha(\rho)$ of $\rho$ in \eqref{eq:LagrangianBW}.
The value of the fixpoint is a zero of the quadratic beta function of $\alpha$ \cite{Pomoni:2009joh}.
It is however not needed for any of the computations performed in the paper, 
since no ultraviolet divergences appear in asymptotic vacuum diagrams of the brick wall theory \eqref{eq:LagrangianBW}.

\section{Star-triangle relation for scalars and fermions}
\label{sec:STRForScalarsAndFermions}
In this section, we first review the star-triangle relation and its consequences for bosons and fermions
\cite{Chicherin:2012yn,Derkachov:2021rrf,Preti:2018vog}. 
This sets the notation and points out parallels to integrable two-dimensional lattice models. For the latter, see also the interesting early work \cite{Au-Yang:1999kid}.
We then proceed to derive the x-unity relation, which is a manipulation of a particular Feynman graph. 
This is the key ingredient needed in later chapters to establish the inversion relations.

In a Wick-rotated $D$-dimensional conformal QFT ($D\in \mathbbm{N}$), the fields depend on coordinates $x^\mu \in \mathbbm{R}^D$ of the Euclidean spacetime. 
For the norm of a vector we write $\left| x \right| = \sqrt{ x^2 } = \sqrt{ (x^1)^2 + \cdots + (x^D)^2}$ and we use the shorthand $x_{ij} := x_i - x_j$. 
In a planar QFT we can make use of integrable lattice model techniques as follows. The planar Feynman diagrams form the lattice. Each vertex carries an infinite dimensional function space, where the ``spin components'' are not labelled by a discrete index, but by the coordinates $x^\mu$. Every summation over the discrete index is replaced by an integration over $x^\mu$. The lattice model's weights are assigned to the edges connecting the vertices, and given by the
generalized coordinate-space propagators 
\begin{equation}
W_u^\ell(x_{10}) 
~:=~ 
\frac{1}{\left( x_{10}^2 \right)^u}
\left[\frac{\slashed{x}_{10}}{\left|x_{10}\right|}\right]^{2\ell}
~=~\adjincludegraphics[valign=c,scale=1]{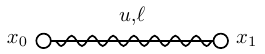}
\label{eq:WeightDefinition}
\end{equation} 
as in \cite{Zamolodchikov:1980mb,Bazhanov:2016ajm}.
They depend on a spectral parameter $u \in \mathbbm{C}$, which eventually is to be tuned to the conformal dimension of the field under consideration, thereby recovering the QFT's standard massless propagator of that field.
Concretely, we have $u=\frac{D-2}{2}$ for scalars and $u=\frac{D-1}{2}$ for fermions. 
The label $\ell \in \left\lbrace 0, \frac{1}{2} \right\rbrace$ is an a-priori undetermined spin label, whose physical values are $\ell = 0$ for scalars and $\ell =\frac{1}{2}$ for fermions.
Inside a Feynman graph, we will indicate $\ell = 0$ by a straight solid line and $\ell > 0$ by a wavy line. 
If $\ell$ is left undetermined, we will draw both on top of each other, as done above. 
In even dimensions, we have $\slashed{x}=\sigma_\mu x^\mu$ for chiral fermions and $\slashed{\bar x}=\bar{\sigma}_\mu x^\mu$ for anti-chiral ones,
where the $\sigma_\mu$,  $\bar \sigma_\mu$ are now\footnote{In the preceding chapter we have already used the notation $\sigma_\mu$ for the special case $D=4$, which should not cause any confusion.} the standard $(2^{D/2-1})$-dimensional $\sigma$-matrices for massless chiral and anti-chiral spinors in $D$ dimensions, satisfying the $D$-dimensional Clifford algebra.
However, in all the diagrams considered in this work the two chiralities always appear in alternation.
Therefore, we will not distinguish them graphically.
For odd dimensions, the slash denotes contraction with the $2^{(D-1)/2}$-dimensional gamma matrices, once again satisfying the associated $D$-dimensional Clifford algebra, i.\ e.\ we write $\slashed{x}=\gamma_\mu x^\mu$.
Finally, the propagators in the theories considered here correspond to complex scalars and complex fermions, which are usually denoted with an arrow on their propagators.
For the sake of simplicity, we will drop these arrows, too.

The spin structure also plays a role when one considers the so called merging rules (see e.\ g.\ \cite{Preti:2018vog}). The spectral parameters of two propagators with coinciding spin $\ell$ and the same initial and end points get added, which translates into
\begin{equation}
W_{u}^\ell(x)W_{-u}^\ell(x)
~=~ 
\mathbbm{I}^{(\ell)}\, ,
\label{eq:Inversion2Weights}
\end{equation}
where the unit matrix $\mathbbm{I}^{(\ell)}$ is the one appropriate to the spin structure on the right-hand side of the $D$-dimensional Clifford algebra: One has $\mathbbm{I}_{2^{D/2-1}}$ for even $D$ and $\mathbbm{I}_{2^{(D-1)/2}}$ for odd $D$ if $\ell = \frac{1}{2}$. 
In the scalar case $\ell = 0$, the unit matrix becomes a scalar factor, $\mathbbm{I}^{(0)} = 1$.

The following calculations are all based on integrability, since we employ the {\it star-triangle relation} (STR) for bosons and fermions \cite{UniquenessDIKazakov,Chicherin:2012yn} (see also \cite{Preti:2018vog,Derkachov:2021rrf} for a review and generalizations). 
Expressed in terms of the generalized propagators $W_u^\ell(x_{jk})$ in \eqref{eq:WeightDefinition}, it reads
\begin{equation}
\begin{split}
\int \dd^{2\eta} x_0  ~~ &
W_{u_1}^0(x_{10}) W_{u_2}^\ell(x_{20}) W_{u_3}^\ell(x_{30})\\
& \stackrel{\sum_i u_i = 2\eta}{~=~} 
r_\ell(u_1,u_2,u_3)\cdot
W_{\eta - u_1}^0(x_{23}) W_{\eta - u_2}^\ell(x_{13}) W_{\eta - u_3}^\ell(x_{12}),
\end{split}
\label{eq:STRgenFormula}
\end{equation}
where we identified the dimension $D$ with twice the so-called crossing parameter $\eta$ \cite{Bazhanov:2016ajm}, i.e.\
\begin{equation}
\eta 
~=~
\frac{D}{2}\; .
\end{equation}
It is straightforward to check that \eqref{eq:STRgenFormula} is invariant under $D$-dimensional conformal transformations: 
Invariance under translations and rotations is manifest. Scale invariance is ensured through the constraint $\sum_i u_i = u_1\!+\!u_2\!+\!u_3=2\eta$. Finally, under inversion $x_\mu \rightarrow x_\mu / x^2$, one has 
$x_{jk}^2 \rightarrow x_{jk}^2 / (x_j^2 x_k^2)$ and $\dd^{2\eta} x_0 \rightarrow \dd^{2\eta} x_0 / (x_0^2)^{2 \eta}$, which also leaves \eqref{eq:STRgenFormula} invariant if and only if the constraint holds. In terms of Feynman diagrams, \eqref{eq:STRgenFormula} becomes
\begin{equation}
\adjincludegraphics[valign=c,scale=0.9]{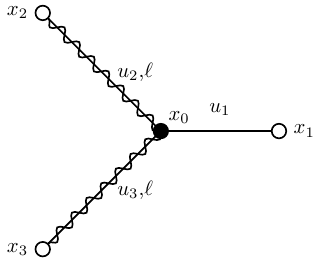} 
\stackrel{\sum_i u_i = 2\eta}{=}
r_\ell(u_1,u_2,u_3)\cdot
\adjincludegraphics[valign=c,scale=0.9]{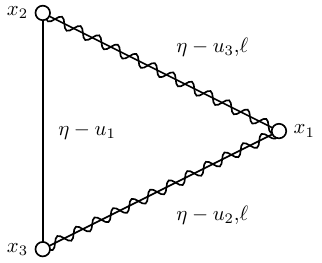} \; , 
\label{eq:STRgenDiag}
\end{equation}
fully justifying the name ``star-triangle relation''. 
Here and below, we use filled, black circles within Feynman diagrams as internal integrated vertices, whereas unfilled circles indicate fixed external points.
Notice the crucial normalization factor
\begin{subequations}
\begin{align}
r_\ell(u_1,u_2,u_3)
~ &:= ~
\pi^{\eta} \cdot
a_0( u_1 )\, a_\ell( u_2 )\, a_\ell( u_3 ) \label{eq:Factor_STR_QFT}\\
\mathrm{with}~~
a_\ell( u ) ~ &:= ~ \frac{\Gamma ( \eta - u + \ell )}{\Gamma ( u + \ell )} \; ,\label{eq:Abbr_alu}
\end{align}\label{eq:Factors_rl_al}%
\end{subequations}
where one easily observes that $a_\ell( u )a_\ell( \eta - u ) = 1$. From the STR \eqref{eq:STRgenFormula} one may derive important lemmas, which we will then use to derive the inversion relations.
\begin{itemize}
\item 
First, thanks to the just discussed conformal covariance of the STR \eqref{eq:STRgenFormula}, one may derive so-called chain relations by sending one of the three external points $x_i\rightarrow \infty$. 
One easily sees that essentially three different limits exist, resulting in three distinct chain rules.
While the resulting three ``chain integrals'' are evaluated in position space, they are formally equivalent to bubble integrals in a dual momentum space interpretation. One finds
\begin{subequations}
\begin{align}
\int \dd^{2\eta} x_0  ~~ 
W_{u_2}^0(x_{20})W_{u_3}^0(x_{30}) 
~=~&
r_0(2\eta-u_2-u_3,u_2,u_3)\cdot
W_{u_2+u_3-\eta}^0(x_{23}) , \\
\adjincludegraphics[valign=c,scale=0.65]{pictures/basics/chainrelations/scalarscalar/scalarscalarLHS.pdf}
~=~&
r_0(2\eta-u_2-u_3,u_2,u_3)\cdot
\adjincludegraphics[valign=c,scale=0.65]{pictures/basics/chainrelations/scalarscalar/scalarscalarRHS.pdf} \; ,
\end{align}\label{eq:ChainRuleEll0}%
\end{subequations}
\begin{subequations}
\begin{align}
\int \dd^{2\eta} x_0  ~~ 
W_{u_1}^0(x_{10}) W_{u_2}^{\frac{1}{2}}(x_{02}) 
~=~& r_{\frac{1}{2}}(u_1,u_2,2\eta-u_1-u_2)\cdot W_{u_1+u_2-\eta}^{\frac{1}{2}}(x_{12}) , \\
\adjincludegraphics[valign=c,scale=0.65]{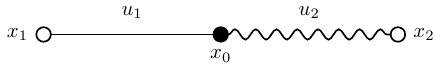} 
~=~&
r_{\frac{1}{2}}(u_1,u_2,2\eta-u_1-u_2)\cdot
\adjincludegraphics[valign=c,scale=0.65]{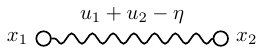} \; ,
\end{align}\label{eq:ChainRuleEll0half}%
\end{subequations}
\begin{subequations}
\begin{align}
\int \dd^{2\eta} x_0  ~~ 
W_{u_2}^{\frac{1}{2}}(x_{20})W_{u_3}^{\frac{1}{2}}(x_{03}) 
\;&=
-\, r_{\frac{1}{2}}(2\eta-u_2-u_3,u_2,u_3)
\cdot \mathbbm{I}^{(\frac{1}{2})}
\cdot W_{u_2+u_3-\eta}^0(x_{23}), \\
\adjincludegraphics[valign=c,scale=0.6]{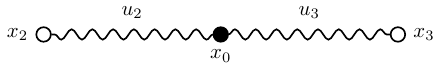}
~&=\,
-\, r_{\frac{1}{2}}(2\eta-u_2-u_3,u_2,u_3)\cdot \mathbbm{I}^{(\frac{1}{2})}\cdot
\adjincludegraphics[valign=c,scale=0.5]{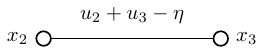} \; .
\end{align}\label{eq:ChainRuleEllHalf}%
\end{subequations}

\item 
Second, by setting the spectral parameter of the scalar line to $u_1 = \eta -\varepsilon$, the scalar side of the triangle on the RHS of \eqref{eq:STRgenFormula} effectively disappears and on the LHS of the remaining equation \eqref{eq:STRgenFormula} the weight $W_{u_1}^0(x_{10})$ divided by $r_\ell(u_1,u_2,u_3)$ acts like killing the integral in the limit $\varepsilon \rightarrow 0$. 
Accordingly, one interprets this combination as a representation of the $D$-dimensional delta function
\begin{equation}
\delta^{(2\eta)}\left( x_{10} \right) 
~=~
\lim_{\varepsilon \rightarrow 0} ~ \pi^{-\eta} a_0 (\varepsilon)
\cdot W_{\eta -\varepsilon}^0(x_{10})
~=~
\adjincludegraphics[valign=c,scale=0.7]{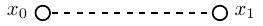},
\label{eq:DeltaDefi}
\end{equation}
which is denoted by a dotted line in a Feynman graph.
Furthermore, observe that setting $u_2 = \eta - u -\varepsilon$ and $u_3 = \eta + u$ in \eqref{eq:ChainRuleEll0} and \eqref{eq:ChainRuleEllHalf}, subsequently multiplying both sides with $\pi^D a_\ell(u+\varepsilon)\, a_\ell (-u)$ and then taking the limit $\varepsilon \rightarrow 0$ gives the prescription of the delta-distribution \eqref{eq:DeltaDefi} on the RHS. 
Here and in the following derivations, we find it very convenient to define\footnote{We introduced the factor $\pi^{-2u}$ for convenience. It drops out in the product $f_\ell (u) f_\ell (-u)$, but turns out to be useful below when constructing the relevant solution of the functional inversion relations \eqref{eq:InversionRelationsKappa}.}
\begin{equation}
f_\ell (u) 
~:=~ 
\pi^\eta 
\pi^{-2u}
a_\ell (u)^{-1}
~=~
\pi^\eta
\pi^{-2u}
\, \frac{\Gamma ( u + \ell )}{\Gamma ( \eta - u + \ell )}.
\label{eq:FactorInversion}
\end{equation}
We may then compactly write for $\ell = 0$
\begin{equation}
\lim_{\varepsilon\rightarrow 0} 
\int \dd^{2\eta} x_0  ~~ 
W_{\eta - u - \varepsilon}^0(x_{20})W_{\eta + u}^0(x_{30}) 
~=~
f_0(u) f_0(-u)\cdot
\delta^{(2\eta)} (x_{23}).
\label{eq:Inversion1Weights}
\end{equation}
\end{itemize}


We now employ the STR and the chain relations to establish useful identities for certain x-shaped Feynman diagrams which we will need later in section \ref{sec:InversionRelationsOfTheFreeEnergyByIntegrabilty}:  
\begin{subequations}
\begin{align}
&\adjincludegraphics[valign=c,scale=0.6]{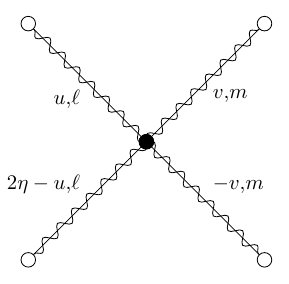}
~&=~
(-1)^{\delta_{m,\frac{1}{2}}}
\cdot
f_\ell (\eta - u)f_\ell (u - \eta)
\cdot ~
&\adjincludegraphics[valign=c,scale=0.6]{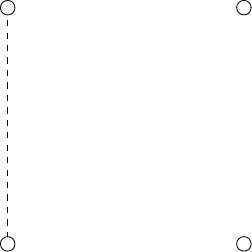} 
\cdot\mathbbm{I}^{(\ell)} \mathbbm{I}^{(m)}\; , \label{eq:XUnity1}\\
&\adjincludegraphics[valign=c,scale=0.6]{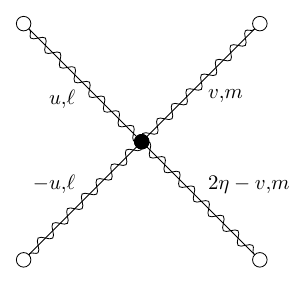}
~&=~
(-1)^{\delta_{l,\frac{1}{2}}}
\cdot
f_m (\eta - v)f_m (v - \eta)
\cdot ~
&\adjincludegraphics[valign=c,scale=0.6]{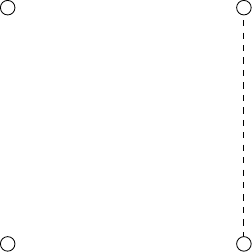}
\cdot\mathbbm{I}^{(\ell)} \mathbbm{I}^{(m)}\,.
\label{eq:XUnity2}
\end{align}\label{eq:XUnity}%
\end{subequations}
We will call them \textit{x-unity} relations and derive them in appendix \ref{app:DerivationOfXUnity}.
They are closely related to the so-called exchange relations \cite{Derkachov:2020zvv}.
The second relation \eqref{eq:XUnity2} may be obtained by relabeling the external points of \eqref{eq:XUnity1}. 


%
%
%
%

\section{Vacuum graphs in the thermodynamic limit}
\label{sec:VacuumGraphsInTheThermodynamicLimit}

\begin{figure}[h]
\includegraphics[scale=0.4]{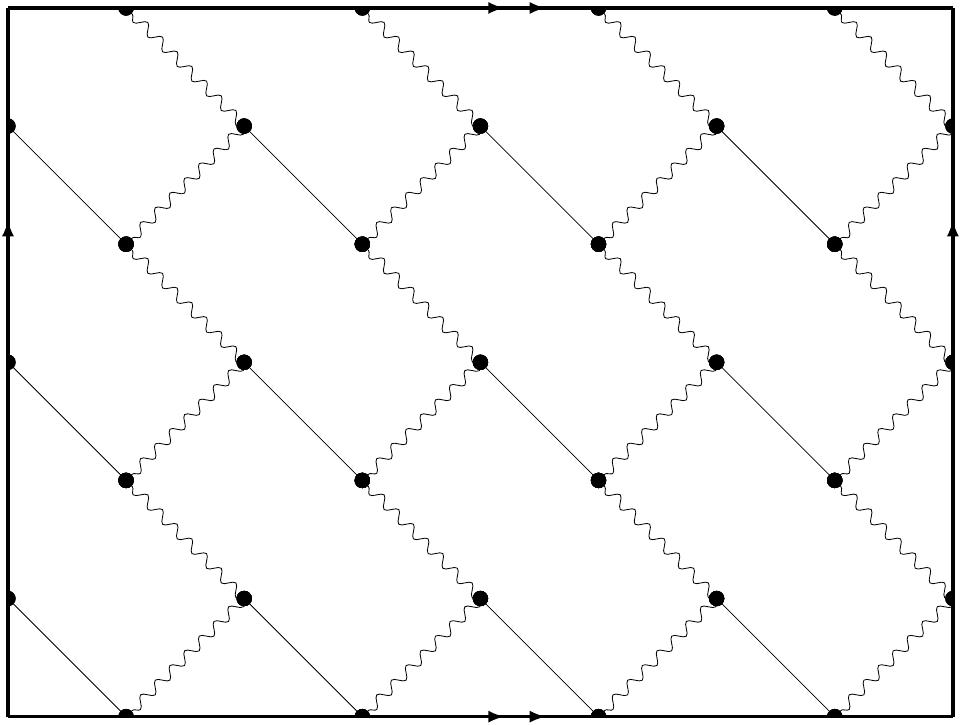}\centering
\caption{The toroidal vacuum graph $Z^{\mathrm{BW}}_{3,4}$ is shown as an example for a contribution to the brick wall model's free energy. It contains three rows, each consisting of four graph building kernels \eqref{eq:GraphBuilderBW}. The faces of the graph generate the typical brick wall pattern of this theory. In the limit $N\rightarrow\infty$, the leading order results in toroidal diagrams. This is impemented into the figure: 
top and bottom lines are identified, and so are the left and right boundary lines.}
\label{fig:BWExample}
\end{figure}

The vacuum diagrams of the brick wall theory form a regular lattice.
One easily checks that, due to the chirality of the vertices $\mathrm{N}\rho\cdot \tr\left( \psi_1\phi \psi_2 \right)$ and $\mathrm{N}\rho\cdot \tr\left( \bar{\psi}_1\phi^\dagger \bar{\psi}_2 \right)$,  no diagrams of spherical topology, i.e.\ ${\cal O}(\mathrm{N}^2)$ in the large-N expansion, exist. Accordingly, the expansion of the free energy, which is a generating function of all of the model's vacuum graphs, starts at $\mathrm{N}^0$, corresponding to graphs of toroidal topology. Thus, this is the leading contribution in the planar limit $\mathrm{N}\rightarrow \infty$ that we consider.

Next, we would like to explain the appearance of the brick wall shape, referring to figure \ref{fig:BWExample} as an illustrating example of a vacuum graph.
The vertices change the flavor of the fermions, therefore no two vertices of the same kind can be adjacent. 
As a consequence, the two flavors of fermions form sub-graphs, which wrap one cycle of the torus.
A generic vacuum diagram $Z^{\mathrm{BW}}_{MN}$ contains $N$ vertical fermion lines, each of them consisting of $M$ $\psi_1$ and $M$ $\psi_2$ propagators. 
The bosonic propagators then connect the fermionic ones such that the faces of the Feynman graphs are hexagonal. 
This gives the characteristic brick wall shape of the diagram.

The regular pattern of the vacuum graphs relies on the fact that it can be generated by a fundamental piece of the diagram, the graph building kernel \cite{Kazakov:2018gcy}
\begin{equation}
\mathbbm{R}^{\mathrm{BW}}
~=~
\adjincludegraphics[valign=c,scale=0.6]{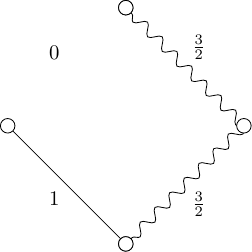}\; . 
\label{eq:GraphBuilderBW}
\end{equation}
It should be understood as an integration kernel.
In order to form the graph $Z^{\mathrm{BW}}_{MN}$, we chain together $N$ graph building kernels to form the row-matrix 
\begin{equation}
\begin{split}
\mathbbm{T}_{N}^{\mathrm{BW}}
~&=~
\tr
\overbrace{
\left[
\mathbbm{R}^{\mathrm{BW}}
\circ \cdots \circ 
\mathbbm{R}^{\mathrm{BW}}
\right]
}^{N-\mathrm{times}} \\
~&=~
\adjincludegraphics[valign=c,scale=0.4]{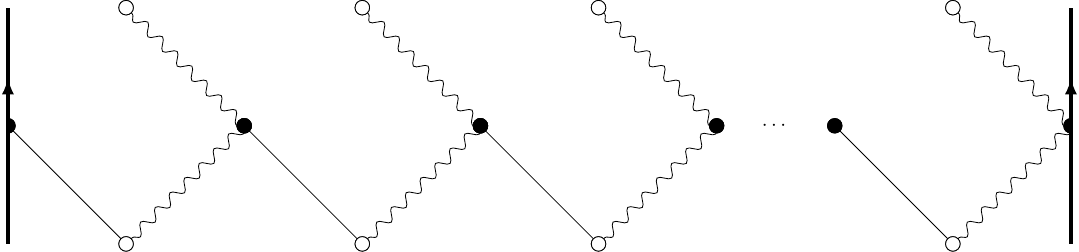} \; .
\end{split}
\end{equation}
Here, the thick black lines should be identified horizontally, as we are computing a trace.
The final step to obtain a brick wall vacuum graph is to convolute $M$ row-matrices on top of each other and to take a vertical trace over the resulting diagram. 
The vertical trace also includes the indices of the elements of the Clifford algebra. 
It yields
\begin{equation}
\begin{split}
& Z^{\mathrm{BW}}_{MN}
~=~
\Tr
\left[
\left(
\mathbbm{T}_{N}^{\mathrm{BW}}
\right)^M
\right] \\
& ~=~
\int
\left[
\prod_{m,n=1}^{M,N} 
\frac{\dd^4 x_{mn}
\dd^4 y_{mn}}{(x_{m,n-1} - y_{mn})^2}
\frac{(x_{mn} - y_{mn})^\mu (\sigma_\mu)^{\alpha_{mn}}_{\dot{\alpha}_{mn}}}{(x_{mn} - y_{mn})^4}
\frac{(x_{mn} - y_{m+1,n})^\nu (\bar{\sigma}_\nu)^{\dot{\alpha}_{mn}}_{\alpha_{m+1,n}}}{(x_{mn} - y_{m+1,n})^4}
\right] ,
\label{eq:ZMNBW}
\end{split}
\end{equation}
a generic vacuum graph of the brick wall theory, see figure \ref{fig:BWExample} for $Z^{\mathrm{BW}}_{3,4}$ as an example. The double-periodic boundary conditions are built into \eqref{eq:ZMNBW}  by the index identifications $M+1\equiv 1$ and $0\equiv N $. 


The free energy is then the infinite sum of all vacuum diagrams
\begin{equation}
Z
~=~
\sum_{M,N=1}^\infty
Z^{\mathrm{BW}}_{MN}
\rho^{2MN} \; ,
\label{eq:VacuumFreeEnergyBW}
\end{equation}
where the two in the exponent of $\rho$ comes from the fact that $\mathbbm{R}^{\mathrm{BW}}$ has four vertices, each shared with one other graph building kernel. 
Note that this exponent might differ for graphs of other theories if some of the weights in their graph building kernel are required to turn into delta functions, see appendix \ref{app:DerivationOfScalarResults}.
The critical coupling $\rho_\mathrm{cr}$ of the brick wall model is defined as the radius of convergence of the series \eqref{eq:VacuumFreeEnergyBW}, which we find to be $\rho_\mathrm{cr}  = \left( \mathbbm{K}^{\mathrm{BW}} \right)^{-\frac{1}{2}}$ with the quantity
\begin{equation}
\mathbbm{K}^{\mathrm{BW}}
~:=~
\lim_{M,N\rightarrow \infty} 
\left|
Z^{\mathrm{BW}}_{MN}
\right|^{\frac{1}{MN}} \; .
\label{eq:FreeEnergyBW}
\end{equation}
It happens to be equal to the free energy in the thermodynamic limit of the brick wall vacuum diagrams $Z^{\mathrm{BW}}_{MN}$. 
This is the quantity we will calculate with the help of integrability by using the STR \eqref{eq:STRgenFormula}.

To proceed, we first generalize the relevant graphs by using the more general graph building kernel 
\begin{equation}
\mathbbm{R}_{\ell m}
\left(\begin{smallmatrix}u_+ & v_+ \\ u_- & v_- \end{smallmatrix}\right)
~=~
\adjincludegraphics[valign=c,scale=0.6]{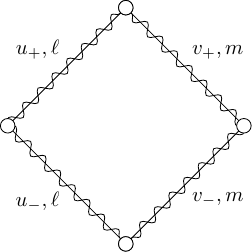} \; .
\label{eq:GraphBuildergen}
\end{equation}
The propagators are the ones in \eqref{eq:WeightDefinition}, and their weights are $u_+$, $u_-$, $v_-$ and $v_+$ (going counterclockwise, starting at the top right propagator) and $\ell$ and $m$ for the ones on the left and on the right, respectively. 
All of them need to be finetuned in order to obtain the desired vacuum graphs of the theory under consideration. 
For the brick wall model this means $\left(\begin{smallmatrix}u_+ & v_+ \\ u_- & v_- \end{smallmatrix}\right) = \left(\begin{smallmatrix} 0 & \frac{3}{2} \\ 1 & \frac{3}{2} \end{smallmatrix}\right)$ and $\ell = 0$, as well as $m = \frac{1}{2}$. 

In the literature \cite{Chicherin:2012yn,Derkachov:2020zvv,Derkachov:2021rrf}, the parameters $u_+$, $u_-$, $v_+$ and $v_-$ of the graph building kernel \eqref{eq:GraphBuildergen} are to be restricted due to the requirement of it being a R-matrix as an element of $\mathrm{End}(V_{\Delta,\ell,\dot{\ell}}\otimes V_{\Delta,\ell,\dot{\ell}})$ of the principle series representation $V_{\Delta,\ell,\dot{\ell}}$ of the Euclidean conformal group.
However, this specification is not necessary in what follows. This is the reason why we neither use the term \textit{R-matrix} for \eqref{eq:GraphBuildergen} nor \textit{transfer matrix} for the row-matrices.

The graph building kernel gets composed to a row-matrix
\begin{equation}
\begin{split}
\mathbbm{T}_{N\;\ell m}
\left(\begin{smallmatrix}u_+ & v_+ \\ u_- & v_- \end{smallmatrix}\right)
~&=~
\tr
\overbrace{
\left[
\mathbbm{R}_{\ell m}
\left(\begin{smallmatrix}u_+ & v_+ \\ u_- & v_- \end{smallmatrix}\right)
\circ \cdots \circ 
\mathbbm{R}_{\ell m}
\left(\begin{smallmatrix}u_+ & v_+ \\ u_- & v_- \end{smallmatrix}\right)
\right]
}^{N-\mathrm{times}} \\
~&=~
\adjincludegraphics[valign=c,scale=0.4]{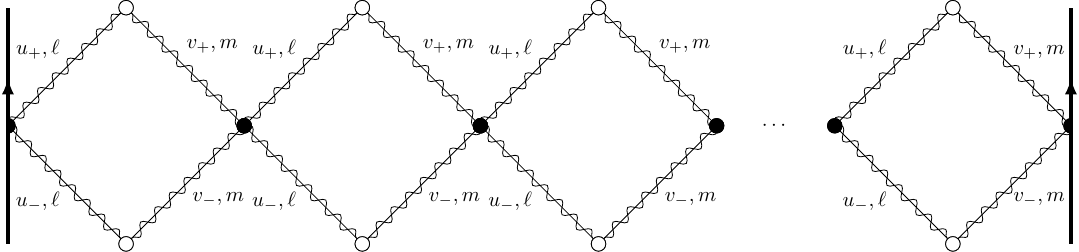} \; , 
\end{split}
\label{eq:TransferMatrixGen}
\end{equation}
such that the generalized graphs are again obtained by stacking up $M$ rows and identifying the vertical cycle of the torus. This is denoted by taking the vertical trace, including any potential spinorial indices.
We find
\begin{equation}
Z_{MN\;\ell m}
\left(\begin{smallmatrix}u_+ & v_+ \\ u_- & v_- \end{smallmatrix}\right)
~=~
\Tr
\left[
\mathbbm{T}_{N\;\ell m}
\left(\begin{smallmatrix}u_+ & v_+ \\ u_- & v_- \end{smallmatrix}\right)^M
\right] \; ,
\end{equation}
and similar to \eqref{eq:FreeEnergyBW}, the generalized graph's free energy in the thermodynamic limit is 
\begin{equation}
\mathbbm{K}_{\ell m}
\left(\begin{smallmatrix}u_+ & v_+ \\ u_- & v_- \end{smallmatrix}\right)
~=~
\lim_{M,N\rightarrow \infty} 
\left|
Z_{MN\;\ell m}
\left(\begin{smallmatrix}u_+ & v_+ \\ u_- & v_- \end{smallmatrix}\right)
\right|^{\frac{1}{MN}} \; .
\label{eq:FreeEnergyGen}
\end{equation}
Finally, we assume that the fact that the trace of a matrix equals the sum of its eigenvalues also holds for the integral operators considered here. 
In the limit $M\rightarrow\infty$ all but the highest eigenvalue are suppressed, so we can write the free energy in the thermodynamic limit as
\begin{equation}
\mathbbm{K}_{\ell m}
\left(\begin{smallmatrix}u_+ & v_+ \\ u_- & v_- \end{smallmatrix}\right)
~=~
\lim_{N\rightarrow \infty}
\left|
\Lambda_{\mathrm{max},N\;\ell m}
\left(\begin{smallmatrix}u_+ & v_+ \\ u_- & v_- \end{smallmatrix}\right)
\right|^{\frac{1}{N}}\; .
\label{eq:FreeEnergyGenHighestEV}
\end{equation}
We have denoted the highest eigenvalue of $\mathbbm{T}_{N\;\ell m}
\left(\begin{smallmatrix}u_+ & v_+ \\ u_- & v_- \end{smallmatrix}\right)$ as $\Lambda_{\mathrm{max},N\;\ell m}
\left(\begin{smallmatrix}u_+ & v_+ \\ u_- & v_- \end{smallmatrix}\right)$. 
The quantity $\mathbbm{K}_{\ell m}
\left(\begin{smallmatrix}u_+ & v_+ \\ u_- & v_- \end{smallmatrix}\right)$ will be derived in the next section in generality, and then specified to the brick wall model. 

\section{Inversion relations of the free energy by integrability}
\label{sec:InversionRelationsOfTheFreeEnergyByIntegrabilty}

The strategy for deriving the inversion relations for the free energy \eqref{eq:FreeEnergyGen} is based on finding an integration kernel that trivializes the generalized transfer matrix \eqref{eq:TransferMatrixGen} to the identity matrix, up to a convolution factor. 
We will construct such an ``inverse'' by usage of the x-unity relations \eqref{eq:XUnity1} and \eqref{eq:XUnity2}. 
It will turn out that there are four of these kernels, leading to four inversion relations for the free energy. 
They are sufficiently constraining to allow for the construction of the desired solution. 
By tuning the spectral parameters to the ones of the brick wall model, we obtain its free energy in the thermodynamic limit. To check our procedure, we verified that the method also reproduces Zamolodchikov's original result \cite{Zamolodchikov:1980mb} in a transparent fashion, see appendix \ref{app:DerivationOfScalarResults}.

To illustrate the inversion up to a factor, consider the convolution of the transfer matrix with its transpose at a different spectral parameter. As an example, we choose $\left(\begin{smallmatrix} -u_+ & 2\eta - v_+ \\2\eta - u_- & -v_- \end{smallmatrix}\right)$. After applying \eqref{eq:XUnity1} a total of $2N$ times, we get the relation
\begin{subequations}
\begin{align}
\mathbbm{T}&_{N\;\ell m} 
\left(\begin{smallmatrix}u_+ & v_+ \\ u_- & v_- \end{smallmatrix}\right)
\circ
\mathbbm{T}_{N\;\ell m}^\mathrm{T}
\left(\begin{smallmatrix} -u_+ & 2\eta - v_+ \\2\eta - u_- & -v_- \end{smallmatrix}\right)\\
~&\stackrel{\phantom{\eqref{eq:XUnity1}}}{=}~ 
\adjincludegraphics[valign=c,scale=0.5]{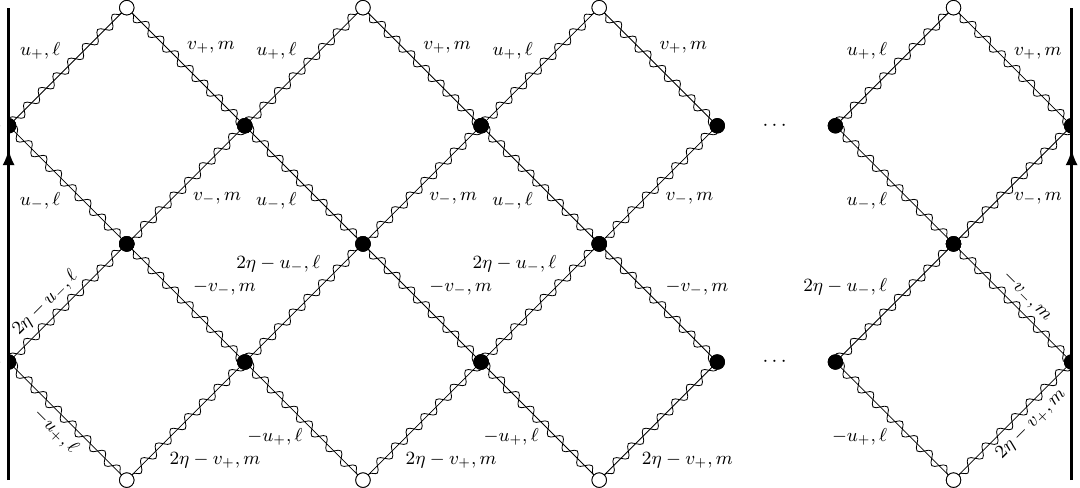}
\label{eq:TransferMatrixAnnihilation1}\\
~&\stackrel{\eqref{eq:XUnity1}}{=}~
\adjincludegraphics[valign=c,scale=0.5]{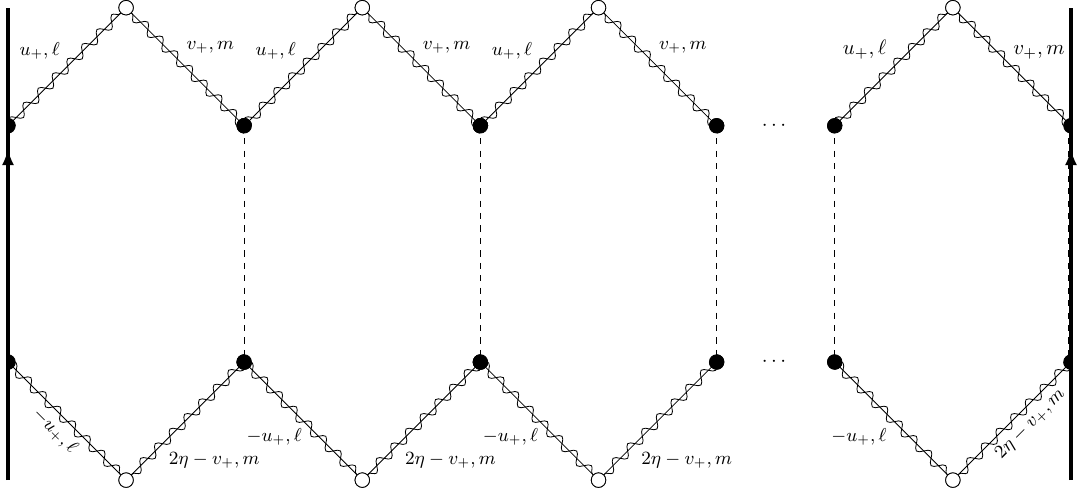} 
\cdot \mathbbm{I}^{(\ell)} \mathbbm{I}^{(m)} \nonumber \\
& \hspace{1.5cm} \cdot 
\left[
(-1)^{\delta_{m,\frac{1}{2}}} f_\ell (\eta - u_-) f_\ell (u_- - \eta)
\right]^N 
\\
~&\stackrel{\phantom{\eqref{eq:XUnity1}}}{=}~
\adjincludegraphics[valign=c,scale=0.5]{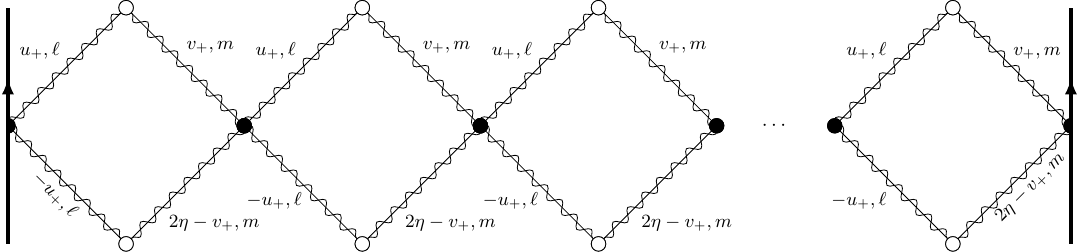}
\cdot \mathbbm{I}^{(\ell)} \mathbbm{I}^{(m)} \nonumber \\
& \hspace{1.5cm} \cdot 
\left[
(-1)^{\delta_{m,\frac{1}{2}}} f_\ell (\eta - u_-) f_\ell (u_- - \eta)
\right]^N
\\
~&\stackrel{\eqref{eq:XUnity1}}{=}~
\adjincludegraphics[valign=c,scale=0.5]{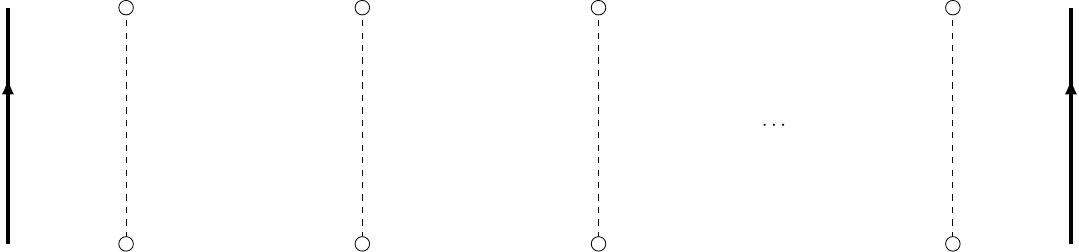} 
\cdot \mathbbm{I}^{(\ell)} \mathbbm{I}^{(m)} \nonumber \\
& \hspace{1.5cm} \cdot
\left[
(-1)^{\delta_{m,\frac{1}{2}}+\delta_{\ell,\frac{1}{2}}} f_\ell (\eta - u_-) f_\ell (u_- - \eta)
f_m (\eta - v_+) f_m (v_+ - \eta)
\right]^N
\\
~&\stackrel{\phantom{\eqref{eq:XUnity1}}}{=}
\left[
(-1)^{\delta_{m,\frac{1}{2}}+\delta_{\ell,\frac{1}{2}}} f_\ell (\eta - u_-) f_\ell (u_- - \eta)
f_m (\eta - v_+) f_m (v_+ - \eta)
\right]^N \cdot \mathbbm{1}_N  \cdot \mathbbm{I}^{(\ell)} \mathbbm{I}^{(m)} 
\label{eq:TransferMatrixAnnihilationEnd}\; .
\end{align}\label{eq:TransferMatrixAnnihilation}%
\end{subequations}
Here, $\mathbbm{1}_N$ is the identity kernel consisting of delta functions.
One could have applied similar steps to the products 
$\mathbbm{T}_{N\;\ell m}
\left(\begin{smallmatrix}u_+ & v_+ \\ u_- & v_- \end{smallmatrix}\right)
\circ
\mathbbm{T}_{N\;\ell m}^\mathrm{T}
\left(\begin{smallmatrix} 2\eta - u_+ &  - v_+ \\ 2\eta - u_- & - v_- \end{smallmatrix}\right)$,
$\mathbbm{T}_{N\;\ell m}
\left(\begin{smallmatrix}u_+ & v_+ \\ u_- & v_- \end{smallmatrix}\right)
\circ
\mathbbm{T}_{N\;\ell m}^\mathrm{T}
\left(\begin{smallmatrix} - u_+ & 2\eta - v_+ \\ - u_- & 2\eta - v_- \end{smallmatrix}\right)$ and
$\mathbbm{T}_{N\;\ell m}
\left(\begin{smallmatrix}u_+ & v_+ \\ u_- & v_- \end{smallmatrix}\right)
\circ
\mathbbm{T}_{N\;\ell m}^\mathrm{T}
\left(\begin{smallmatrix} 2\eta - u_+ & - v_+ \\ - u_- & 2\eta - v_- \end{smallmatrix}\right)$.
Consequently, one would have used \eqref{eq:XUnity1} and \eqref{eq:XUnity2}, \eqref{eq:XUnity2} and \eqref{eq:XUnity1} and \eqref{eq:XUnity2} and \eqref{eq:XUnity2}. In total, this leads to four relations similar to \eqref{eq:TransferMatrixAnnihilation}. We proceed the same way in all four cases. We let the product of transfer matrices act on the eigenvector corresponding to the highest eigenvalue, take their magnitude to the power $\frac{1}{N}$ and take the limit $N\rightarrow \infty$ as in \eqref{eq:FreeEnergyGenHighestEV}. Since the eigenvalues are insensitive to transposition, the LHS of \eqref{eq:TransferMatrixAnnihilation1} turns into a product of free energy contributions at different spectral parameters. The RHS \eqref{eq:TransferMatrixAnnihilationEnd} yields its scalar factor by these steps. Applying these manipulations to the four relations yields the four inversion relations for the free energy $\mathbbm{K}_{\ell m}
\left(\begin{smallmatrix} u_+ & v_+ \\ u_- & v_- \end{smallmatrix}\right)$:
\begin{subequations}
\begin{align}
&\mathbbm{K}_{\ell m}
\left(\begin{smallmatrix} u_+ & v_+ \\ u_- & v_- \end{smallmatrix}\right)
\mathbbm{K}_{\ell m}
\left(\begin{smallmatrix} - u_+ & 2\eta - v_+ \\ 2\eta - u_- & - v_- \end{smallmatrix}\right)
~&&=~
f_\ell (\eta - u_-) f_\ell (u_- - \eta)
f_m (\eta - v_+) f_m (v_+ - \eta) , \\
&\mathbbm{K}_{\ell m}
\left(\begin{smallmatrix} u_+ & v_+ \\ u_- & v_- \end{smallmatrix}\right)
\mathbbm{K}_{\ell m}
\left(\begin{smallmatrix} 2\eta - u_+ &  - v_+ \\ 2\eta - u_- & - v_- \end{smallmatrix}\right)
~&&=~
f_\ell (\eta - u_-) f_\ell (u_- - \eta)
f_\ell (\eta - u_+) f_\ell (u_+ - \eta) , \\
&\mathbbm{K}_{\ell m}
\left(\begin{smallmatrix} u_+ & v_+ \\ u_- & v_- \end{smallmatrix}\right)
\mathbbm{K}_{\ell m}
\left(\begin{smallmatrix} - u_+ & 2\eta - v_+ \\ - u_- & 2\eta - v_- \end{smallmatrix}\right)
~&&=~
f_m (\eta - v_-) f_m (v_- - \eta)
f_m (\eta - v_+) f_m (v_+ - \eta) , \\
&\mathbbm{K}_{\ell m}
\left(\begin{smallmatrix} u_+ & v_+ \\ u_- & v_- \end{smallmatrix}\right)
\mathbbm{K}_{\ell m}
\left(\begin{smallmatrix} 2\eta - u_+ & - v_+ \\ - u_- & 2\eta - v_- \end{smallmatrix}\right)
~&&=~
f_m (\eta - v_-) f_m (v_- - \eta)
f_\ell (\eta - u_+) f_\ell (u_+ - \eta) .
\end{align} \label{eq:InversionRelationsFreeEnergy}%
\end{subequations}
Next, to construct a solution for \eqref{eq:InversionRelationsFreeEnergy}, we make the ansatz
\begin{equation}
\mathbbm{K}_{\ell m}
\left(\begin{smallmatrix} u_+ & v_+ \\ u_- & v_- \end{smallmatrix}\right)
~=~
\kappa^1_{\ell}(u_+)\kappa^2_{\ell}(u_-)\kappa^3_{m}(v_+)\kappa^4_{m}(v_-)
\label{eq:FreeEnergyAnsatz}
\end{equation}
with a-priori four functions $\kappa^1_\ell (u)$, $\kappa^2_\ell (u)$, $\kappa^3_\ell (u)$ and $\kappa^4_\ell (u)$ yet to be determined. Plugging the ansatz \eqref{eq:FreeEnergyAnsatz} into the four inversion relations \eqref{eq:InversionRelationsFreeEnergy}, we find that all four functions are identical: 
$\kappa^1_\ell (u) = \kappa^2_\ell (u) = \kappa^3_\ell (u) = \kappa^4_\ell (u) = :\kappa_\ell (u)$. It has to satisfy the functional relations
\begin{subequations}
\begin{align}
\kappa_\ell (u) \kappa_\ell (-u) ~&=~ 1\, ,\label{eq:InversionRelationsKappa1}\\
\frac{\kappa_\ell (\eta - u)}{\kappa_\ell (u)} ~&=~ f_\ell (u)\; ,\label{eq:InversionRelationsKappa2}
\end{align}\label{eq:InversionRelationsKappa}%
\end{subequations}
in slight generalization of a result in \cite{Bazhanov:2016ajm}.
We can construct a solution satisfying \eqref{eq:InversionRelationsKappa} by using the explicit form of $f_\ell (u)$ and alternately requiring the function $\kappa_\ell (u)$ to satisfy \eqref{eq:InversionRelationsKappa1} and \eqref{eq:InversionRelationsKappa2} \cite{Bombardelli:2016scq,Shankar:1977cm,Zamolodchikov:1976uc}. 
This yields the solution
\begin{equation}
\kappa_\ell (u) 
~=~
\pi^u
\frac{\Gamma (\eta - u + \ell)}{\Gamma (\eta + \ell)}
\prod_{k=1}^\infty
\frac{\Gamma (2\eta k + \eta - u + \ell) \, \Gamma (2\eta k + u + \ell) \, \Gamma (2\eta k - \eta + \ell)}
{\Gamma (2\eta k - \eta + u + \ell) \, \Gamma (2\eta k - u + \ell) \, \Gamma (2\eta k + \eta + \ell)} \; ,
\label{eq:Kappagenfinalresult}
\end{equation}
which reduces to the findings of \cite{Bazhanov:2016ajm} after setting $\ell = 0$.
We may now apply the general solution 
\begin{equation}
\mathbbm{K}_{\ell m}
\left(\begin{smallmatrix} u_+ & v_+ \\ u_- & v_- \end{smallmatrix}\right)
~=~
\kappa_{\ell}(u_+)\kappa_{\ell}(u_-)\kappa_{m}(v_+)\kappa_{m}(v_-)
\label{eq:KKgenfinalresult}
\end{equation}
to the brick wall graphs by setting the spectral parameters to $\left(\begin{smallmatrix} 0 & \frac{3}{2} \\ 1 & \frac{3}{2} \end{smallmatrix}\right)$ and $\ell = 0$, as well as $m = \frac{1}{2}$. 
Putting everything together, the free energy of the brick wall graphs reads
\begin{equation}
\begin{split}
\mathbbm{K}^{\mathrm{BW}}
~&=~
\mathbbm{K}_{0,\frac{1}{2}}
\left(\begin{smallmatrix} 0 & \frac{3}{2} \\ 1 & \frac{3}{2} \end{smallmatrix}\right)
~=~
\kappa_{0}(0)\kappa_{0}(1)\kappa_{\frac{1}{2}}\left(\tfrac{3}{2}\right)\kappa_{\frac{1}{2}}\left(\tfrac{3}{2}\right)\\
~&\stackrel{2\eta=D=4}{=}~
1\cdot \pi^{\frac{3}{2}} \frac{\Gamma \left( \frac{5}{4} \right)}{\Gamma \left( \frac{3}{4} \right)} \cdot \frac{\pi^2}{2} \cdot \frac{\pi^2}{2}
~=~
\frac{\pi^{\frac{11}{2}}}{16}
\frac{\Gamma \left( \frac{1}{4} \right)}{\Gamma \left( \frac{3}{4} \right)} \; .
\end{split}
\label{eq:finalresult}
\end{equation}
The resulting critical coupling of the brick wall theory \eqref{eq:LagrangianBW} is then $\rho_\mathrm{cr}  = \left[ \frac{\pi^{\frac{11}{2}}}{16}
\frac{\Gamma \left( \frac{1}{4} \right)}{\Gamma \left( \frac{3}{4} \right)} \right]^{-\frac{1}{2}}$. 

Finally, we make the important observation that $\mathbbm{K}$ factorizes into contributions $\kappa$ of the individual propagators contained in the graph building kernel, see \eqref{eq:KKgenfinalresult}.
The connectivity of the vacuum diagrams depends on how many of the propagators in the graph building kernel are either tuned such that they vanish or else turn into a delta function. 
In both cases the contribution to the free energy is trivial, i.e.\  $\kappa_\ell (0) = 1$, $\lim_{\varepsilon \rightarrow 0} 
\pi^{-\eta} a_0 (\varepsilon)\kappa_0 (\eta - \varepsilon) = 1$ (cf.\ appendix \ref{app:DerivationOfScalarResults}).
Thus, in the thermodynamic limit, the topology of the vacuum graphs is solely reflected in the exponent of the free energy.
However, its value is sensitive to the different values of $u$ and $\ell$ in the evaluation of \eqref{eq:Kappagenfinalresult}.
For $\ell = 0$ and $u = \frac{D-2}{2}$, i.e.\ for a bosonic field in $D$ dimensions, we find the compact result
\begin{equation}
\kappa_0 \left( \tfrac{D-2}{2} \right)
~\stackrel{2\eta = D}{=}~
\frac{\pi ^{\frac{D-1}{2}} \Gamma \left(\frac{3}{2}-\frac{1}{D}\right)}{\Gamma \left(\frac{D}{2}\right) \Gamma \left(1-\frac{1}{D}\right)}\; .
\end{equation}
For the fermionic case $\ell = \frac{1}{2}$ and $u = \frac{D-1}{2}$, we unfortunately were not able to deduce a similarly compact form.
Accordingly, here we only present some specific results for $\kappa_\frac{1}{2} \left( \tfrac{D-1}{2} \right)$: for, respectively, $D=3,4,6$ one has $\frac{2\pi}{\sqrt{3}}$, $\frac{\pi^2}{2}$ and $\frac{\pi^3}{2\sqrt{6}}$.

\section{Conclusions and outlook}
\label{sec:Outlook}
Let us summarize our work. Our focus is on large vacuum energy diagrams of the brick wall model, whose Lagrangian is given in \eqref{eq:LagrangianBW} with \eqref{eq:doubletraceBW}. This is an integrable, planar, chiral and non-unitary four-dimensional matrix quantum field theory, and was first proposed in \cite{Pittelli:2019ceq}.
It may be considered an integrable relative of the standard non-integrable Yukawa theory, in close analogy with bi-scalar fishnet theory, which in the same vein may be considered to be an integrable relative of standard non-integrable $\phi^4_4$-theory. The typical Feynman-diagrams of this bi-fermionic model are of regular ``brick-wall''-type, replacing the regular square lattices of standard fishnet theory. Curiously, in contradistinction to the fishnet theory, and more generally to the $\chi_0$- and $\chi$-CFTs, the model may not be obtained as a double-scaling limit of planar, twisted $\mathcal{N}=4$ SYM; instead, one instead needs to consider twisted $\mathcal{N}=2$ SCFT \cite{Pittelli:2019ceq}. Our key result is then the adaptation of  A.\ B.\ Zamolodchikov's powerful computation of the thermodynamic free energy of fishnet graphs to the brick-wall case. This allows us to find this quantity in the closed form \eqref{eq:finalresult}.

Our graphical view of the brick wall model immediately suggests the formulation of two further planar QFTs, thereby continuing the quest for integrable QFTs that are not derivable from deformations of maximally supersymmetric $\mathcal{N}=4$ SYM.
A similar motivation drove, perhaps, the recent construction of the ``loom'' models of \cite{Kazakov:2022dbd}. 
First, replacing the fermions of the brick wall model \eqref{eq:LagrangianBW} by two different bosons and thus making it a tri-scalar theory does not fundamentally change the diagrammatics in the limit $\mathrm{N}\rightarrow \infty$. 
The vacuum Feynman diagrams are still of brick wall, or, equivalently, honeycomb form.
However, now the interaction terms are exactly marginal in six as opposed to four dimensions.
This leads us to write down the following non-unitary, but integrable cousin of $\phi^3_6$-theory with Lagrangian 
\begin{equation}
\begin{split}
&\mathcal{L}^{\phi^3}_{D=6} 
~=~
\mathrm{N}  \cdot \mathrm{tr}\left[ 
\frac{1}{2}\sum_{j=1}^3 \partial^\mu \phi_j^\dagger \partial_\mu \phi_j^{ }
+ 
\rho
\cdot
\left(
\phi_1\phi_2 \phi_3 + \phi^\dagger_1\phi_2^\dagger \phi_3^\dagger
\right)
\right] \; .
\end{split}
\label{eq:LagrangianPhi3}
\end{equation}
It is equivalent to one of the models formulated in \cite{Mamroud:2017uyz}, where in addition the anomalous dimension of this theory's length-two operators was calculated in a non-perturbative fashion.
Clearly this theory generates honeycomb graphs with propagators living in six dimensions. Here it is totally unclear whether the model is derivable from a unitary supersymmetric strongly twisted SYM theory, as no suitable candidates are known in dimension six.
However, the corresponding free energy in the thermodynamic limit was also calculated as the third of the three cases considered in \cite{Zamolodchikov:1980mb}. 
As a check, we successfully reproduce his result by our methods in appendix \ref{app:DerivationOfScalarResults} in \eqref{eq:FreeEnergyHoneyComb}.
Second, our ad-hoc graphical view of the brick wall model \eqref{eq:LagrangianBW}, inspired (but not derived) from deformed $\mathcal{N}=4$ SYM may also be applied to the $\gamma$-deformation of ABJM theory \cite{Caetano:2016ydc}.
Keeping only the boson-fermion interactions in the analogue of the $\chi_0$-theory \eqref{eq:LagrangianChi0} then yields the following three-dimensional Lagrangian
\begin{equation}
\begin{split}
&\mathcal{L}^{\mathrm{ABJM~FN}}_{D=3}
~=~
-\mathrm{N}  \cdot \mathrm{tr}\left[ 
\sum_{j=1}^2 \partial^\mu Y_j^\dagger \partial_\mu Y_j^{\phantom{\dagger}}
+ \I \sum_{k=1}^2 \Psi^\dagger_k \slashed{\partial} \Psi_{k}^{\phantom{\dagger}}
+
\rho 
\cdot 
\left(
Y_1^{\phantom{\dagger}} Y_2^\dagger \Psi_1^{\phantom{\dagger}} \Psi_2^\dagger
-Y_1^\dagger Y_2^{\phantom{\dagger}} \Psi_1^\dagger \Psi_2^{\phantom{\dagger}}
\right)
\right] \; .
\end{split}
\label{eq:LagrangianABJMFN}
\end{equation}
In contrast to the four-dimensional case, the vacuum diagrams do not follow a brick wall pattern but have the appearance of square-lattice fishnets.
The difference to the vacuum graphs of the bi-scalar fishnet theory \eqref{eq:LagrangianFN} is that now there are zig-zag lines formed by fermionic propagators in alternation with bosonic zig-zag lines. 
Once again, we are able to model these graphs by suitably composing the generalized graph building kernel \eqref{eq:GraphBuildergen}, evaluated as 
$\mathbbm{R}_{0 ,\frac{1}{2}}
\left(\begin{smallmatrix}\frac{1}{2} & 1 \\ \frac{1}{2} & 1 \end{smallmatrix}\right)$.
Tracing out the steps of section \ref{sec:InversionRelationsOfTheFreeEnergyByIntegrabilty}, we find the free energy of the theory \eqref{eq:LagrangianABJMFN} in the thermodynamic limit to be 
\begin{equation}
\mathbbm{K}_{0 ,\frac{1}{2}}
\left(\begin{smallmatrix}\frac{1}{2} & 1 \\ \frac{1}{2} & 1 \end{smallmatrix}\right)
~=~
\kappa_0 (\tfrac{1}{2})^2  \kappa_\frac{1}{2} (1)^2
~\stackrel{2\eta = D =3}{=}~
\frac{4\pi^3}{27}
\frac{\Gamma \left(\frac{1}{6}\right)^2}{\Gamma \left(\frac{2}{3}\right)^2} \; .
\end{equation} 
Clearly it would again be interesting to see whether the model \eqref{eq:LagrangianABJMFN} may be derived as a double scaling limit from a supersymmetric strongly twisted SCFT in three dimensions.

The brick-wall model \eqref{eq:LagrangianBW}, as well as the theories \eqref{eq:LagrangianPhi3} and \eqref{eq:LagrangianABJMFN},
should allow for an adaption of the calculations of a large number of quantities that have been computed exactly for the fishnet model. For the brick wall model, some of these have already been performed in \cite{Pittelli:2019ceq}, such as the leading order of $\alpha(\rho)$ in \eqref{eq:LagrangianBW} and the anomalous dimension of the operator $\mathrm{tr}\left[ \phi^L \right]$. 
The diagonalization of the brick wall graph building kernel \eqref{eq:GraphBuilderBW} was already examined in \cite{Derkachov:2020zvv} and would allow to get a closed form for Basso-Dixon \cite{Basso:2017jwq} brick wall diagrams.
Conceptually, the construction of a ``brickchain'' similar to the fishchain, a quantum mechanical model based on the graph building kernel \cite{Gromov:2019aku,iakhibbaev2023generalising}, would give hints on a dual AdS description of the brick wall model.
The fact that the brick wall theory cannot be obtained from a double-scaling limit of $\mathcal{N}=4$ SYM makes a holographic dual description very intriguing.

The method of inversion relations does not yet account for twisted toroidal diagrams, where the fermion lines wrap the cycle of the torus multiple times before closing. 
It would be interesting to implement them in the framework, however the result for the fishnet theory should not be altered, since it was confirmed by TBA \cite{Basso:2018agi}.
A confirmation by TBA for the brick wall result would be an equally good sanity check.
Finally, the brick wall model, similar to the fishnet model, should serve as a stepping stone to uncover the integrable structure of more evolved theories.
For this reason, translating methods from those two simplest toy models to the $\chi_0$-, $\chi$-CFT and eventually to $\mathcal{N}=4$ SYM is of high priority.

\section*{Acknowledgments}
We thank Matthias Volk for initial collaboration on this project.
We are thankful to Changrim Ahn, Zoltan Bajnok and Lance Dixon for very useful discussions. Special thanks to Lance Dixon, Enrico Olivucci and Volodya Kazakov for very careful readings of our manuskript, and several valuable suggestions on references.
This work is funded by the Deutsche Forschungsgemeinschaft (DFG, German Research Foundation)
- Projektnummer 417533893/GRK2575 “Rethinking Quantum Field Theory”.
This research was supported in part by grant NSF PHY-1748958 to the Kavli Institute for Theoretical Physics (KITP).

\appendix
\section{Derivation of x-unity}
\label{app:DerivationOfXUnity}
In this appendix, we will prove the auxiliary relations stated in \eqref{eq:XUnity} by using the STR and the chain rules. 
In detail, the calculation goes as follows:
\begin{subequations}
\begin{align}
&\adjincludegraphics[valign=c,scale=0.57]{pictures/xmove/left/LHS/xmove1.pdf}
~=~
\lim_{\varepsilon\rightarrow 0}
\adjincludegraphics[valign=c,scale=0.57]{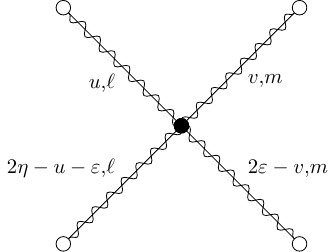} \\
~&\stackrel{\eqref{eq:Inversion1Weights}}{=}~
\lim_{\varepsilon\rightarrow 0}
\adjincludegraphics[valign=c,scale=0.57]{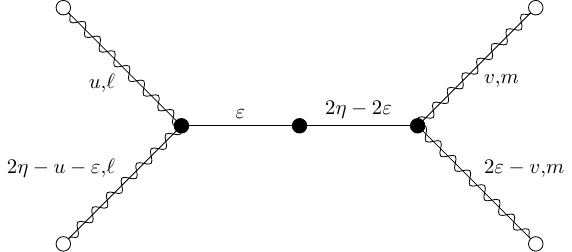}
\cdot
\frac{1}{f_0(\eta - 2\varepsilon) f_0(2\varepsilon - \eta)} \\
~&\stackrel{\eqref{eq:STRgenFormula}}{=}~
\lim_{\varepsilon\rightarrow 0}
\adjincludegraphics[valign=c,scale=0.57]{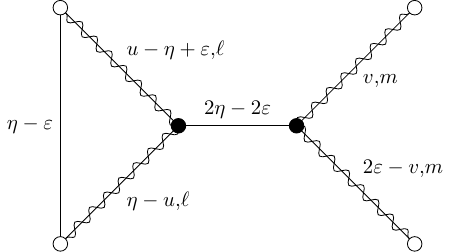}
\cdot
\frac{ r_\ell (\varepsilon , u , 2\eta - u - \varepsilon ) }{f_0(\eta - 2\varepsilon) f_0(2\varepsilon - \eta)} \\
~&\stackrel{\eqref{eq:DeltaDefi},\eqref{eq:STRgenFormula}}{=}~
\lim_{\varepsilon\rightarrow 0}
\adjincludegraphics[valign=c,scale=0.57]{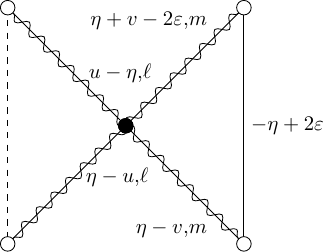}
\cdot
\frac{ r_\ell (\varepsilon , u , 2\eta - u - \varepsilon ) }{f_0(\eta - 2\varepsilon) f_0(2\varepsilon - \eta)} 
\frac{\pi^\eta}{a_0(\varepsilon)}
r_m (2\eta - 2\varepsilon , v , 2\varepsilon - v ) \\
~&\stackrel{\eqref{eq:Inversion2Weights}}{=}~
\adjincludegraphics[valign=c,scale=0.57]{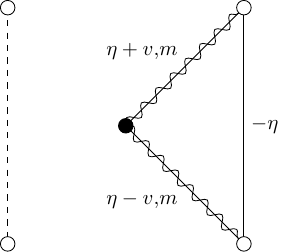}
\cdot
\lim_{\varepsilon\rightarrow 0}
\frac{ r_\ell (\varepsilon , u , 2\eta - u - \varepsilon ) }{f_0(\eta - 2\varepsilon) f_0(2\varepsilon - \eta)} 
\frac{\pi^\eta}{a_0(\varepsilon)}
r_m (2\eta - 2\varepsilon , v , 2\varepsilon - v )
\cdot \mathbbm{I}^{(\ell)}\\
~&\stackrel{\eqref{eq:ChainRuleEllHalf}\; \mathrm{or}\; \eqref{eq:ChainRuleEll0}}{=}~
\adjincludegraphics[valign=c,scale=0.57]{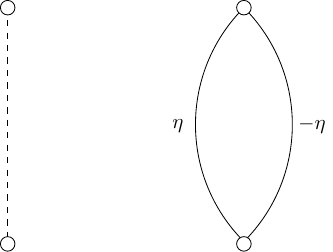}
\cdot
(-1)^{\delta_{m,\frac{1}{2}}} r_m (0 , \eta + v , \eta -v )\nonumber \\
& \hspace{2cm} 
\cdot\lim_{\varepsilon\rightarrow 0}
\frac{ r_\ell (\varepsilon , u , 2\eta - u - \varepsilon ) }{f_0(\eta - 2\varepsilon) f_0(2\varepsilon - \eta)} 
\frac{\pi^\eta}{a_0(\varepsilon)}
r_m (2\eta - 2\varepsilon , v , 2\varepsilon - v )
\cdot \mathbbm{I}^{(\ell)} \mathbbm{I}^{(m)}\\
~&\stackrel{\eqref{eq:Inversion2Weights}}{=}~
(-1)^{\delta_{m,\frac{1}{2}}}
\cdot
f_\ell (\eta - u)f_\ell (u - \eta)
\cdot
\adjincludegraphics[valign=c,scale=0.57]{pictures/xmove/left/RHS/xmove7.pdf} ~
\cdot \mathbbm{I}^{(\ell)} \mathbbm{I}^{(m)}
\end{align}%
\end{subequations}
In the last step, we simplified the overall factor by extensive use of $a_m(u) a_m (\eta - u)  = 1$ and the definition of $f_\ell(u)$ in \eqref{eq:FactorInversion}.

\section{Derivation of scalar results}
\label{app:DerivationOfScalarResults}
In this appendix, we will carefully re-derive Zamolodchikov's expression for the free energy per propagator \cite[eq.(18)]{Zamolodchikov:1980mb} from our general solution \eqref{eq:KKgenfinalresult} and \eqref{eq:Kappagenfinalresult}. This appears to be a worthwhile exercise, given the extremely concise arguments and derivations in \cite{Zamolodchikov:1980mb}.

To begin, we restrict ourself to scalar propagator weights in this section. 
Furthermore, we notice the different normalization of the weights $G_\alpha(x)$ in \cite[eq.(8)]{Zamolodchikov:1980mb} compared to the ones considered here, $W_u^0(x)$ , in \eqref{eq:WeightDefinition}. We observe that we can relate them as 
\begin{equation}
W_{u}^0(x)
~=~
\pi^{u}
\frac{\Gamma (\eta - u)}{\Gamma (\eta)}
G_{\pi - \pi \frac{u}{\eta}}(x)
~=:~
h(u)
G_{\pi - \pi \frac{u}{\eta}}(x) \; .
\label{eq:WWeightToGWeight}
\end{equation}
The graph building kernel in terms of the weight $G_\alpha(x)$, $\mathbbm{R}^{G}$, is related to the generalized one \eqref{eq:GraphBuildergen} by
\begin{equation}
\mathbbm{R}_{0,0}
\left(\begin{smallmatrix}u_+ & v_+ \\ u_- & v_- \end{smallmatrix}\right)
~=~
h\left(\eta - \eta \tfrac{\alpha_+}{\pi}\right)
h\left(\eta - \eta \tfrac{\alpha_-}{\pi}\right)
h\left(\eta - \eta \tfrac{\beta_+}{\pi}\right)
h\left(\eta - \eta \tfrac{\beta_-}{\pi}\right)
\cdot
\mathbbm{R}^G
\left(\begin{smallmatrix}\alpha_+ & \beta_+ \\ \alpha_- & \beta_- \end{smallmatrix}\right)
\end{equation}
and we introduced Zamolodchikov's angles on the medial lattice $\frac{\alpha_\pm}{\pi} := 1 - \frac{u_\pm}{\eta}$ and $\frac{\beta_\pm}{\pi} := 1 - \frac{v_\pm}{\eta}$ \cite{Kazakov:2022dbd}. 
According to \eqref{eq:FreeEnergyGen}, the normalized free energy is 
\begin{equation}
\begin{split}
\mathbbm{K}^G 
\left(\begin{smallmatrix}\alpha_+ & \beta_+ \\ \alpha_- & \beta_- \end{smallmatrix}\right)
~&=~
\frac{
\mathbbm{K}_{0,0}
\left(\begin{smallmatrix}\eta - \eta \frac{\alpha_+}{\pi} & \eta - \eta \frac{\beta_+}{\pi} \\ \eta - \eta \frac{\alpha_-}{\pi} & \eta - \eta \frac{\beta_-}{\pi} \end{smallmatrix}\right)
}{
h\left(\eta - \eta \frac{\alpha_+}{\pi}\right)
h\left(\eta - \eta \frac{\alpha_-}{\pi}\right)
h\left(\eta - \eta \frac{\beta_+}{\pi}\right)
h\left(\eta - \eta \frac{\beta_-}{\pi}\right)
}\\
~&=:~
\kappa^G(\alpha_+)
\kappa^G(\alpha_-)
\kappa^G(\beta_+)
\kappa^G(\beta_-) \; ,
\end{split}\label{eq:FreeEnergyWeightG}
\end{equation} 
which implies for the factors
\begin{equation}
\begin{split}
\kappa^G(\alpha)
~&=~
\frac{\kappa_0 \left( \frac{D}{2} - \frac{D\alpha}{2\pi} \right)}{h\left( \frac{D}{2} - \frac{D\alpha}{2\pi} \right)}\\
~&=~
\prod_{k=1}^\infty
\frac{\Gamma (D k + D/2 -  D\alpha/2\pi) \, \Gamma (D k + D\alpha/2\pi) \, \Gamma (D k - D/2)}
{\Gamma (D k - D/2 + D\alpha/2\pi) \, \Gamma (D k - D\alpha/2\pi) \, \Gamma (2\eta k + D/2)}
~=:~
\frac{1}{Z(\alpha)} \; .
\end{split}
\label{eq:ZOfAlpha}
\end{equation}
$Z(\alpha)$ is Zamolodchikov's result for the free energy per propagator in the thermodynamic limit, so is the factor $\kappa^G(\alpha)$. 
We obtain the inverse of $Z(\alpha)$, since he defines $Z(\alpha):=\e^{-f(\alpha)}$ with $f(\alpha) = \mathrm{log}\, \kappa^G(\alpha)$.

Next, we derive the expressions for the critical coupling of regular triangular, quadratic and honeycomb lattices in, respectively, $D=3,4,6$ dimensions from the general free energy per weight \eqref{eq:ZOfAlpha}, as considered in \cite{Zamolodchikov:1980mb}. 
However, therein the critical coupling was computed in relation to yet another weight, called $G\mathcal{D}(x) := s^{-1}\cdot  G_\frac{2\pi}{D}(x)$ \cite[eq.(1)]{Zamolodchikov:1980mb}, and the factor of proportionality is $s = 2\pi, 4\pi, 8\pi$, respectively to $D=3,4,6$. 
The motivation for the new weight is that $G\mathcal{D}(x)$ is the $D$-dimensional Fourier transform of the scalar momentum space propagator. 
Under the assumption that the Lagrangian contains an interaction term $g\cdot \mathcal{L}_\mathrm{int}$, we will demonstrate how to calculate $g_\mathrm{cr}  = \left( \mathbbm{K}^{G\mathcal{D}} \right)^{-b}$ in the three cases. 
The quantity $b$ denotes the ratio of the number of graph building kernels to the number of vertices in the vacuum diagrams and will be determined below.
\begin{itemize}
\item
In three dimensions, the Feynman graphs can have a triangular shape, which is due to a sixtic interaction, as provided by e.\ g.\ the ABJM fishnet theory \cite{Caetano:2016ydc}. 
In order to obtain a triangular graph building kernel from \eqref{eq:GraphBuildergen}, we have to turn one propagator into a delta-distribution. 
We pick the north-west propagator to undergo the metamorphosis and by \eqref{eq:DeltaDefi} we can construct the triangular graph building kernel as
\begin{equation}
\mathbbm{R}^{D=3}
~:=~
\lim_{\varepsilon \rightarrow 0} 
\pi^{-\eta} a_0 (\varepsilon)
\cdot
\mathbbm{R}_{0,0}
\left(\begin{smallmatrix}\eta - \varepsilon & \frac{1}{2} \\ \frac{1}{2} & \frac{1}{2} \end{smallmatrix}\right)
~=~
\adjincludegraphics[valign=c,scale=0.6]{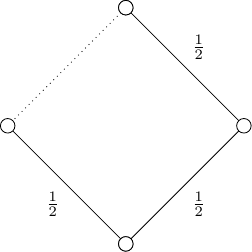} 
\; .
\end{equation}
We observe that, due to the delta-distribution, the number of vertices equals the number of graph building kernels in a asymptotic vacuum diagram, hence $b = 1$.
The corresponding free energy related to the canonical weight function $W_u^0(x)$ is 
\begin{equation}
\begin{split}
\mathbbm{K}^{D=3} 
~:&=~
\lim_{\varepsilon \rightarrow 0} 
\pi^{-\eta} a_0 (\varepsilon)
\cdot
\mathbbm{K}_{0,0}
\left(\begin{smallmatrix}\eta - \varepsilon & \frac{1}{2} \\ \frac{1}{2} & \frac{1}{2} \end{smallmatrix}\right)
~=~
\kappa_0 \left( \tfrac{1}{2} \right)^3
\cdot
\lim_{\varepsilon \rightarrow 0} 
\pi^{-\eta} a_0 (\varepsilon)
\kappa_0 (\eta - \varepsilon)\\
~&=~
\left[
\frac{\sqrt{\pi}}{3}
\frac{\Gamma (\frac{1}{6})}{\Gamma (\frac{2}{3})}
\right]^3 \; ,
\end{split}
\end{equation}
which is obtained by \eqref{eq:KKgenfinalresult} and \eqref{eq:Kappagenfinalresult}. 
Regarding the weight $G_\alpha (x)$, the correcting factor yielding a delta function is provided by $h(u)$ in \eqref{eq:WWeightToGWeight}.
By noting that $u = \eta$ and $u = \frac{1}{2}$ correspond to $\alpha = 0$ and $\alpha = \frac{2\pi}{3}$ for $D = 3$, respectively, one can calculate directly by \eqref{eq:FreeEnergyWeightG} and \eqref{eq:ZOfAlpha}
\begin{equation}
\mathbbm{K}^{G,D=3}
~=~ 
\frac{1}{Z(0) Z(\frac{2\pi}{3})^3}
~=~
\left[
\frac{\sqrt{\pi}}{6}
\frac{\Gamma (\frac{1}{6})}{\Gamma (\frac{2}{3})}
\right]^3 \; .
\end{equation}
Changing to the weight $G\mathcal{D}(x)$ via dividing by $s=2\pi$ for each of the three remaining propagators, see the explanation in the paragraph following \eqref{eq:ZOfAlpha}, gives the critical coupling 
\begin{equation}
g_\mathrm{cr}^{D=3}
~=~
\left( \mathbbm{K}^{G\mathcal{D},D=3} \right)^{-1}
~=~
\left( \frac{\mathbbm{K}^{G,D=3}}{s^3} \right)^{-1}
~=~
\left[
12 \sqrt{\pi}\;
\frac{\Gamma (\frac{2}{3})}{\Gamma (\frac{1}{6})}
\right]^3 
~=~
\frac{1}{c_3}\; .
\end{equation}
This is the original result by Zamolodchikov \cite{Zamolodchikov:1980mb}, which was reproduced by TBA in \cite{Basso:2018agi}.
\item
In four dimensions, the Feynman graphs under consideration have four-valent vertices.
A theory producing them is the bi-scalar fishnet theory \cite{Gurdogan:2015csr}.
The corresponding graph building kernel solely consists of scalar propagators in four dimensions with weight $u = 1$ and takes the form
\begin{equation}
\mathbbm{R}^{D=4}
~:=~
\mathbbm{R}_{0,0}
\left(\begin{smallmatrix}1 & 1 \\ 1 & 1 \end{smallmatrix}\right)
~=~
\adjincludegraphics[valign=c,scale=0.6]{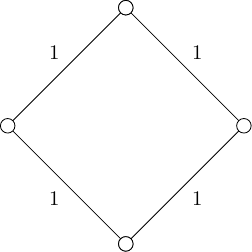}
\; ,
\end{equation}
and the ratio of graph building kernels to vertices is $b = \frac{1}{2}$.
The free energy corresponding to the scalar canonical weight \eqref{eq:WeightDefinition} is 
\begin{equation}
\mathbbm{K}^{D=4} 
~:=~
\mathbbm{K}_{0,0}
\left(\begin{smallmatrix}1 & 1 \\ 1 & 1 \end{smallmatrix}\right)
~=~
\kappa_0 \left( 1 \right)^4
~=~
\left[
\frac{\pi ^{3/2}}{4}
\frac{\Gamma \left(\frac{1}{4}\right)}{\Gamma \left(\frac{3}{4}\right)}
\right]^4 \; .
\end{equation}
Transitioning to the weights $G_\alpha (x)$ by \eqref{eq:FreeEnergyWeightG}, we note that $\alpha = \frac{\pi}{2}$ for $D=4$ such that we find
\begin{equation}
\mathbbm{K}^{G,D=4}
~=~ 
\frac{1}{Z(\frac{\pi}{2})^4}
~=~
\left[
\frac{\sqrt{\pi}}{4}
\frac{\Gamma \left(\frac{1}{4}\right)}{\Gamma \left(\frac{3}{4}\right)}
\right]^4 \; .
\end{equation}
In order to recover Zamolodchikov's famous result \cite{Zamolodchikov:1980mb}, we need to once again properly normalize the weight, see above. 
Choosing $s = 4\pi$, we indeed find the critical coupling to be
\begin{equation}
g_\mathrm{cr}^{D=4}
~=~
\left( \mathbbm{K}^{G\mathcal{D},D=4} \right)^{-\frac{1}{2}}
~=~
\left( \frac{\mathbbm{K}^{G,D=4}}{s^4} \right)^{-\frac{1}{2}}
~=~
\left[
16 \sqrt{\pi}\;
\frac{\Gamma (\frac{3}{4})}{\Gamma (\frac{1}{4})}
\right]^2
~=~
\frac{1}{c_4}\; .
\end{equation}
This result was also obtained by different techniques in \cite{Basso:2018agi}.
\item
The last class of graphs under consideration contain cubic interactions in $D=6$.
We propose the theory \eqref{eq:LagrangianPhi3}, first introduced in \cite{Mamroud:2017uyz}, which has these graphs as their contribution to the free energy. 
Originally, they were coined honeycomb graphs, but essentially they are the scalar versions of the brick wall graphs considered above. 
Consequently, the graph building kernel has the same shape as \eqref{eq:GraphBuilderBW} but with $D=6$ scalar propagators of weight $u = 2$, therefore we write
\begin{equation}
\mathbbm{R}^{D=6}
~:=~
\mathbbm{R}_{0,0}
\left(\begin{smallmatrix}0 & 2 \\ 2 & 2 \end{smallmatrix}\right)
~=~
\adjincludegraphics[valign=c,scale=0.6]{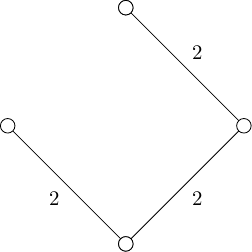}
\; .
\end{equation}
We observe that the ratio $b$ is again $\frac{1}{2}$ and the canonical free energy is found by \eqref{eq:KKgenfinalresult} to be
\begin{equation}
\mathbbm{K}^{D=6} 
~:=~
\mathbbm{K}_{0,0}
\left(\begin{smallmatrix}0 & 2 \\ 2 & 2 \end{smallmatrix}\right)
~=~
\kappa_0 \left( 0 \right)\cdot \kappa_0 \left( 2 \right)^3
~=~
\left[
\frac{\pi ^{5/2}}{6}
\frac{\Gamma \left(\frac{1}{3}\right)}{ \Gamma \left(\frac{5}{6}\right)}
\right]^3 \; .
\label{eq:FreeEnergyHoneyComb}
\end{equation}
The free energy in terms of the weights $G_\alpha (x)$ can be computed via \eqref{eq:WWeightToGWeight} after recalling that the non-zero angle is $\alpha = \frac{\pi}{3}$ in $D=6$. It yields
\begin{equation}
\mathbbm{K}^{G,D=6}
~=~ 
\frac{1}{Z(\pi)Z(\frac{\pi}{3})^3}
~=~
\left[
\frac{\sqrt{\pi}}{3}
\frac{\Gamma \left(\frac{1}{3}\right)}{\Gamma \left(\frac{5}{6}\right)}
\right]^3 \; .
\end{equation}
Finally, we again turn to the weight $G\mathcal{D} (x)$ by dividing by $s=8\pi$ for each non-vanishing propagator in the graph building kernel. 
The critical coupling related to this weight is then found to be
\begin{equation}
g_\mathrm{cr}^{D=6}
~=~
\left( \mathbbm{K}^{G\mathcal{D},D=6} \right)^{-\frac{1}{2}}
~=~
\left( \frac{\mathbbm{K}^{G,D=6}}{s^3} \right)^{-\frac{1}{2}}
~=~
\left[
24 \sqrt{\pi}\;
\frac{\Gamma (\frac{5}{6})}{\Gamma (\frac{1}{3})}
\right]^\frac{3}{2}
~=~
\frac{1}{c_6}\; ,
\end{equation}
which completes the reproduction of the results of \cite{Zamolodchikov:1980mb}.
\end{itemize}

\bibliography{paper_brickwall}
\bibliographystyle{OurBibTeX}
\end{document}